\documentclass[twoside,11pt]{article}

\usepackage{blindtext}

\usepackage{jmlr2e}

\usepackage{amssymb}

\usepackage{booktabs}
\usepackage{xspace}
\usepackage{mathtools}
\usepackage[ruled, vlined]{algorithm2e}
\usepackage{enumitem}
\usepackage{subcaption}
\usepackage{svg}
\usepackage{graphicx}
\usepackage{bbding}
\usepackage{multirow}
\usepackage{amsmath}
\usepackage{breqn}
\usepackage{algorithmic}
\usepackage{listings}
\usepackage{ulem}

\newcommand{\name}{\text{Galvatron}\xspace}

\usepackage{lastpage}

\ShortHeadings{\name: An Automatic Distributed System for Efficient Foundation Model Training}{Liu, Wang, Zhu, Fu, Liu, Lin and Cui}
\firstpageno{1}

\begin{document}

\title{\name: An Automatic Distributed System for Efficient Foundation Model Training}

\author{\name Xinyi Liu\textsuperscript{1*} \email  xy.liu@stu.pku.edu.cn \\
    \name Yujie Wang\textsuperscript{1*} \email alfredwang@pku.edu.cn \\
    \name Shenhan Zhu\textsuperscript{1*} \email shenhan.zhu@pku.edu.cn \\
    \name Fangcheng Fu\textsuperscript{1}\email  ccchengff@pku.edu.cn \\
    \name Qingshuo Liu\textsuperscript{1} \email  liu.qingshuo@stu.pku.edu.cn \\
    \name Guangming Lin\textsuperscript{1} \email linguangming.pku@gmail.com \\
    \name Bin Cui\textsuperscript{1} \email bin.cui@pku.edu.cn \\
    \addr \textsuperscript{1} Key Lab of High Confidence Software Technologies (MOE), School of CS, Peking University, China \\
    \addr \textsuperscript{*} Equal contribution. \\
    \vspace{-3em}
}

\editor{}

\maketitle

\begin{abstract}
\name is a distributed system for efficiently training large-scale Foundation Models. 
It overcomes the complexities of selecting optimal parallelism strategies by automatically identifying the most efficient hybrid strategy, incorporating data, tensor, pipeline, sharded data, and sequence parallelism, along with recomputation.
The system's architecture includes a profiler for hardware and model analysis, a search engine for strategy optimization using decision trees and dynamic programming, and a runtime for executing these strategies efficiently. 
Benchmarking on various clusters demonstrates \name's superior throughput compared to existing frameworks. 
This open-source system offers user-friendly interfaces and comprehensive documentation, making complex distributed training accessible and efficient. The source code of \name is available at \url{https://github.com/PKU-DAIR/Hetu-Galvatron}.
\end{abstract}

\begin{keywords}
  Distributed Training System, Automatic Parallelism, Foundation Models
\end{keywords}

\section{Introduction}
\vspace{-3pt}
In recent years, large-scale Transformer-base Foundation Models, particularly Large Language Models (LLMs), have demonstrated exceptional performance in text understanding and generation~\citep{DBLP:journals/corr/VaswaniSPUJGKP17, DBLP:journals/corr/abs-2303-18223, DBLP:conf/nips/BrownMRSKDNSSAA20, DBLP:journals/corr/abs-2303-08774, DBLP:journals/corr/abs-2312-11805, DBLP:journals/corr/abs-2407-21783}. Scaling laws suggest that increasing model parameters enhances performance~\citep{DBLP:journals/corr/abs-2001-08361}, driving the need for large-scale distributed training. To optimize such training, various parallelization methods have been proposed, including data parallelism~\citep{DBLP:conf/nips/DeanCMCDLMRSTYN12, DBLP:conf/nips/BrownMRSKDNSSAA20}, tensor parallelism~\citep{DBLP:conf/sc/NarayananSCLPKV21}, pipeline parallelism~\citep{DBLP:conf/nips/HuangCBFCCLNLWC19, DBLP:conf/sosp/NarayananHPSDGG19, DBLP:conf/icml/NarayananPSCZ21}, sharded data parallelism~\citep{DBLP:conf/sc/RajbhandariRRH20, DBLP:journals/pvldb/ZhaoGVLHXWSOSDB23}, and sequence/context parallelism~\citep{DBLP:conf/mlsys/KorthikantiCLMA23, DBLP:journals/corr/abs-2309-14509, DBLP:journals/corr/abs-2310-01889}.  Each method presents distinct memory, computation, and communication characteristics.
 
Transformer-based Foundation Models exhibit various architectural designs, with decoder-only models (e.g., GPT, Llama) being predominant for its general ability, while encoder-only and encoder-decoder models like BERT~\citep{DBLP:conf/naacl/DevlinCLT19} and T5~\citep{DBLP:journals/jmlr/RaffelSRLNMZLL20} are widely used in specific downstream applications~\citep{DBLP:journals/corr/abs-2303-18223}.
Within the decoder-only category, notable examples include GPT~\citep{DBLP:conf/nips/BrownMRSKDNSSAA20}, Llama~\citep{DBLP:journals/corr/abs-2407-21783}, and Qwen~\citep{DBLP:journals/corr/abs-2412-15115}, each exhibiting unique design and workload characteristics. Additionally, models of the same type often have multiple configurations, influencing their computational and memory properties.

Beyond architectural differences, training settings introduce additional complexity. For instance, the context length during Llama 3 training is gradually extended from 8K to 128K~\citep{DBLP:journals/corr/abs-2407-21783}, resulting in significant workload variations, even for models with identical configurations. Furthermore, the hardware infrastructure employed, such as cluster setups and hardware specifications, impacts computational speed and bandwidth. These factors collectively pose challenges in designing efficient parallelization strategies tailored to specific workloads.

In existing deep learning systems like DeepSpeed~\citep{DBLP:conf/kdd/RasleyRRH20} and Megatron~\citep{DBLP:conf/sc/NarayananSCLPKV21}, selecting parallel strategies often relies on expert experiences and laborious tuning, requiring substantial time yet often yielding suboptimal results. To address this, we introduce \name, a distributed training system supporting automatic parallelization. Its main features include:

\noindent \textbf{1. Comprehensive and Fine-Grained Parallelism}: \name supports all major parallelization strategies, including data, tensor, pipeline, sharded data, and sequence parallelism, along with recomputation, which trades off between computation overhead and memory consumption. It enables \textit{layer-level} customization, allowing each Transformer layer to adopt independent parallel strategies for better efficiency and adaptability.

\noindent \textbf{2. Efficient Automatic Parallelism Optimization}: \name's search engine evaluates computation, memory, and communication requirements for various strategies. Through dynamic programming, it selects strategies that balance memory usage with computation and communication overhead, adapting to hardware constraints for optimal performance.

\noindent \textbf{3. Versatility Across Architectures and Hardware}: \name is compatible with diverse model architectures, including language and vision models, and supports various hardware platforms such as NVIDIA GPUs, Ascend NPUs~\citep{ascendAI2023}, and Hygon DCUs, making it suitable for a wide range of workloads.
    
\noindent \textbf{4. User-Friendly Design}: \name provides intuitive interfaces, enabling seamless integration with minimal code changes to replace original models with hybrid parallel versions, reducing adoption complexity.
    
\noindent \textbf{5. State-of-the-Art Performance}: With automated parallel optimization, \name achieves superior training throughput compared to other frameworks, ensuring efficient large-scale Transformer model training.

So far, \name has demonstrated superior efficiency compared to existing frameworks and has been adopted as the underlying framework for various works~\citep{DBLP:conf/asplos/WangWZFLXLLW025,DBLP:conf/asplos/WangZFMZZHLC25}. Beyond academic research, \name has also been deployed in industrial applications of prominent companies such as ByteDance, Huawei, ZTE, BAAI, and etc.

\vspace{-7pt}
\section{System Design}
\vspace{-3pt}

In this section, we introduce the system design of \name, which comprises three tightly integrated core modules: the profiler, the search engine, and the runtime. These components collaboratively facilitate efficient distributed training, as illustrated in Fig.~\ref{fig:architecture}.

\begin{minipage}[b]{0.39\textwidth}
  \centering
  \includegraphics[width=\textwidth]{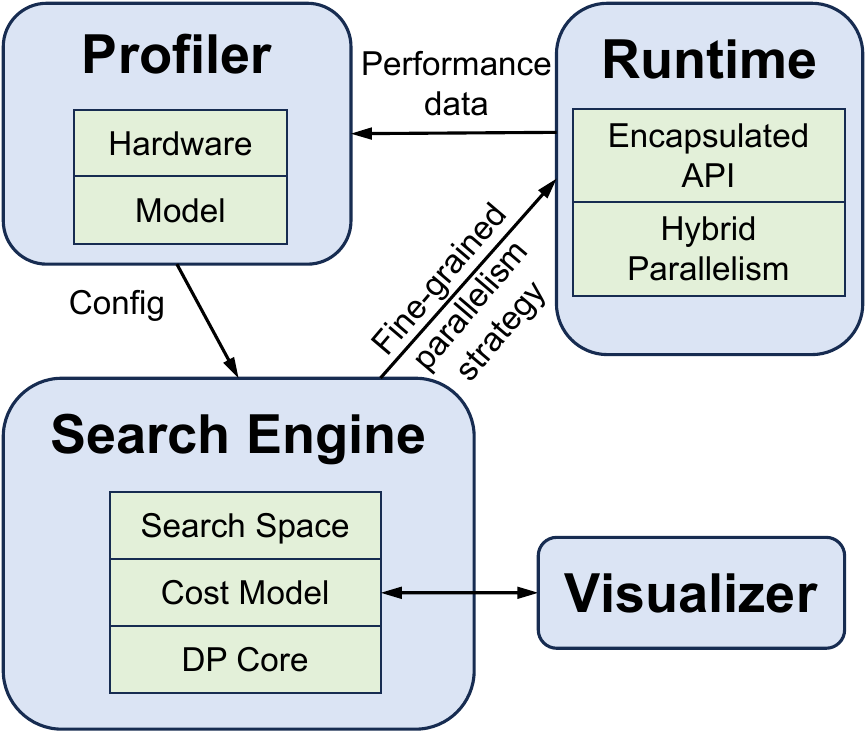}
  \captionof{figure}{\small{System architecture.}}
  \label{fig:architecture}
\end{minipage}
\hfill
\begin{minipage}[b]{0.60\textwidth}
  \centering
  \includegraphics[width=\textwidth]{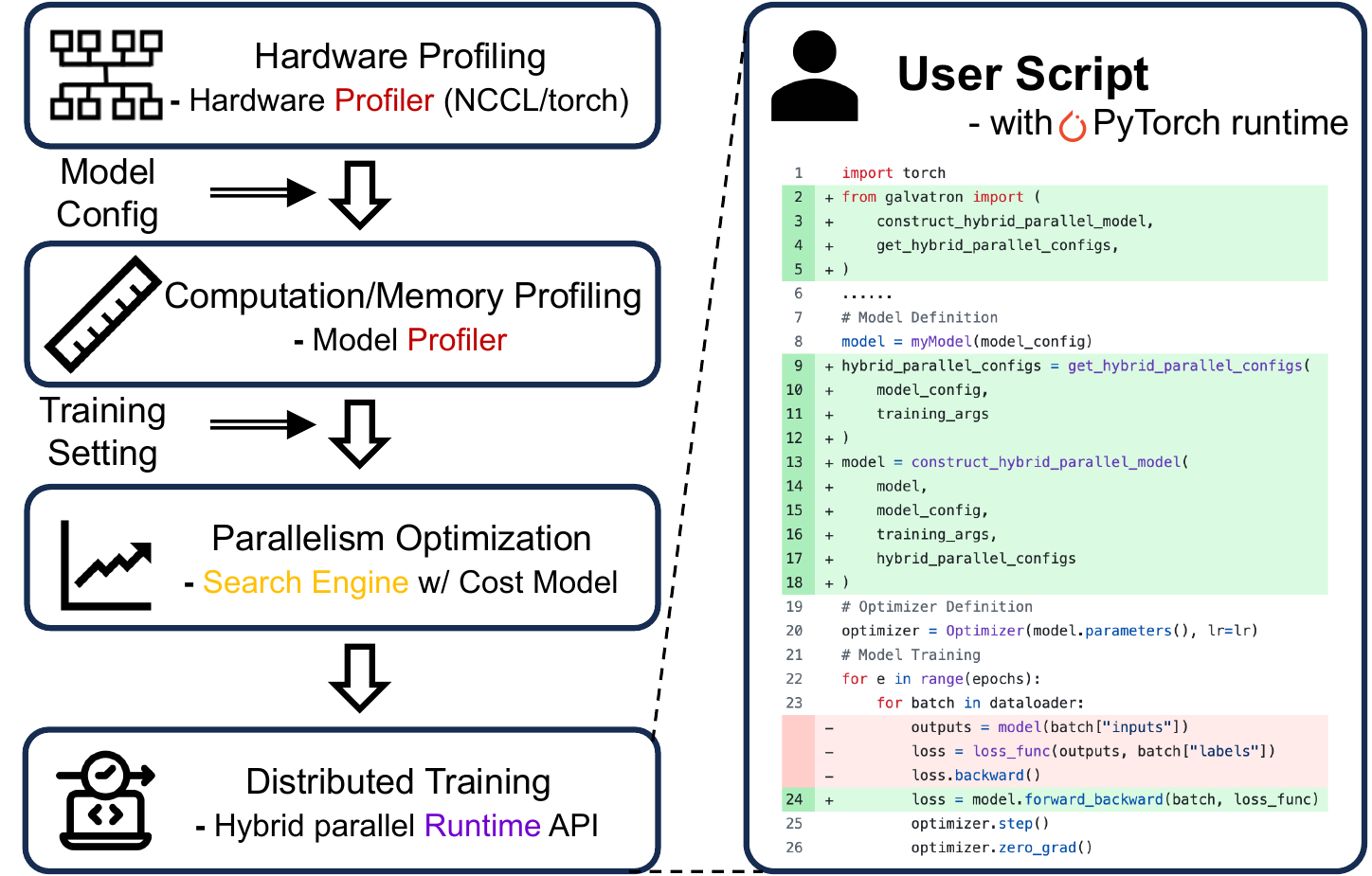}
  \captionof{figure}{\small{System workflow.}}
  \label{fig:workflow}
\end{minipage}
\vspace{3pt}

\noindent \textbf{Profiler}: This module conducts an in-depth analysis of the hardware capabilities and model execution characteristics. It evaluates inter-device communication bandwidth and single-device computational throughput.
For model analysis, it examines computational patterns and memory requirements across components, providing essential insights for constructing precise cost models that form the basis for intelligent strategy selection.

\noindent \textbf{Search Engine}: Serving as the decision-making core of the system, this module leverages profiled data to identify optimal parallelization strategies within minutes. It models the parallel configuration space using decision trees, discards infeasible configurations, and constructs accurate cost models for time and memory across strategies. Through dynamic programming, it automatically determines the most effective combination of parallel strategies for each model layer~\citep{DBLP:journals/pvldb/MiaoWJSNZ022}. Additionally, \name includes a visualization plugin for the cost model, enhancing user accessibility.

\noindent \textbf{Runtime}: This module efficiently implements all major parallel strategies, including data, tensor, pipeline, sharded data, and sequence parallelism, while treating recomputation as a distinct parallel dimension. It encapsulates high-level strategies into efficient hybrid parallel models. As shown in Fig.~\ref{fig:workflow}, the runtime retrieves hybrid parallel configurations via \lstinline{get_hybrid_parallel_configs} (line 9) and replaces the original model with the corresponding strategy using \lstinline{construct_hybrid_parallel_model} (line 13), requiring minimal modifications to the existing project.

This design provides a robust and flexible execution environment that adapts to diverse hardware configurations and model architectures. The seamless integration of these modules achieves the excellent efficiency of distributed training.

\vspace{-7pt}
\section{Workflow}
\vspace{-3pt}
Fig.~\ref{fig:workflow} illustrates the workflow for automated distributed parallel training using \name, divided into four key steps. Firstly, the hardware profiler evaluates the cluster environment, measuring communication bandwidth across devices at different scales.
Secondly, given a user-defined model, the model profiler analyzes the computation time and memory requirements of each layer, including model states and activations. Thirdly, the search engine leverages the profiled data to construct a cost model and determines the optimal fine-grained hybrid parallel strategy. Finally, the runtime module carries out the distributed training process based on the selected strategy.
Users are only required to specify the hardware environment configuration and Transformer model details, while \name autonomously manages the optimization process, from profiling to execution, ensuring both ease of use and high performance. \name simplifies distributed training with user-friendly scripts and comprehensive API documentation for runtime conversion. It strikes a balance between automation and customization, enabling effortless deployment for standard use cases while offering fine-grained control for specialized scenarios.

\vspace{-7pt}
\section{Benchmark}
\vspace{-3pt}
\begin{figure}[!t]
\centering
\includegraphics[width=\textwidth]{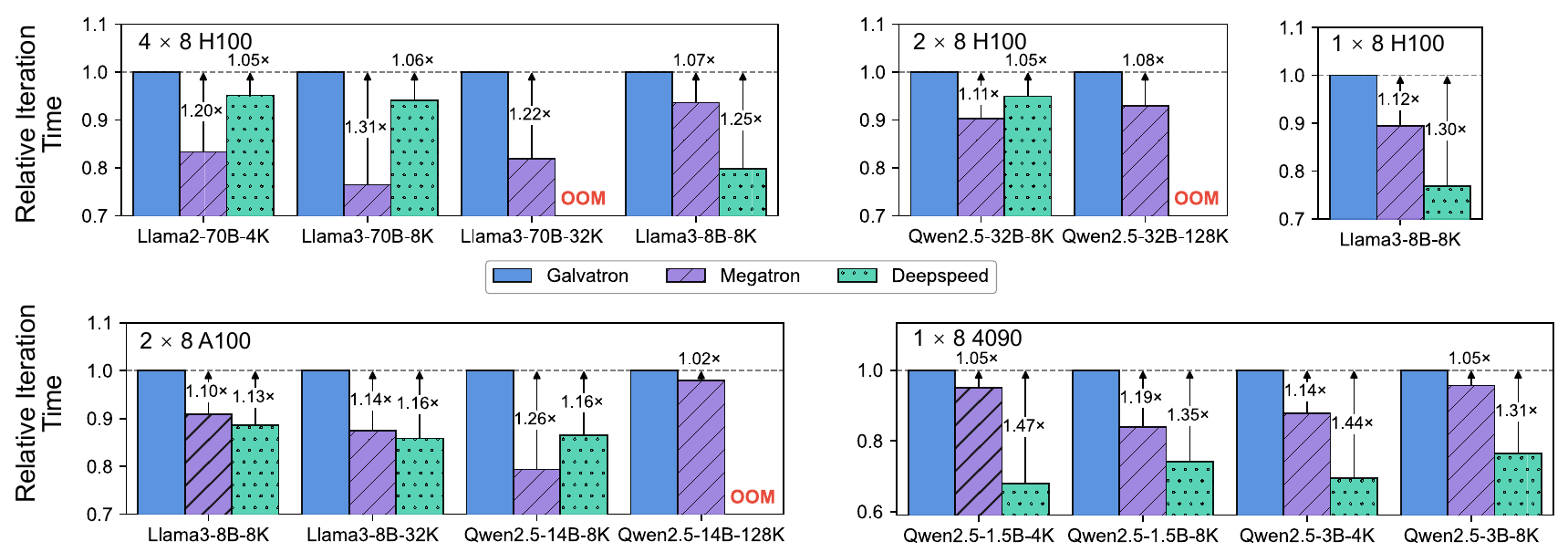}
\caption{End-to-end speedup of different clusters. (OOM indicates that all strategies result in an "out of memory" error.)}
\label{fig:e2e}
\vspace{-20pt}
\end{figure}

To evaluate the effectiveness of \name's fine-grained automatic parallelization, we conduct experiments on clusters equipped with NVIDIA H100, A100, and 4090 GPUs at varying scales. We compare \name with state-of-the-art distributed training frameworks, including Megatron~\citep{DBLP:conf/sc/NarayananSCLPKV21} and DeepSpeed~\citep{DBLP:conf/kdd/RasleyRRH20}, employing manual tuning to determine the optimal parallel strategies. The results presented in Fig.~\ref{fig:e2e} indicate that \name achieves up to 1.26–1.47× higher throughput compared to these frameworks.
Unlike Megatron and DeepSpeed that exhibit specific strengths and limitations across varying cluster and model configurations, \name consistently presents the best efficiency thanks to its automatic adjustment of fine-grained parallelism.

\vspace{-7pt}
\section{Conclusion}
\vspace{-3pt}
This paper introduces \name, an open-source system for fine-grained automatic parallel training of large-scale Foundation Models that autonomously identifies optimal parallel strategies based on varying workloads, while providing user-friendly interfaces. The source code is available at \url{https://github.com/PKU-DAIR/Hetu-Galvatron}. For detailed information, APIs, and more functionalities, please refer to our documentation.\footnote{\url{https://hetu-galvatron.readthedocs.io}}

\vskip 0.2in
\nocite{*}
\bibliography{sample}

\end{document}